\documentstyle[seceq,mbf,wrapft]{ptptex}

\pubinfo{Vol.~95, No.~4, April 1996}  
\notypesetlogo  
\preprintnumber[3cm]{
KUNS-1325\\PTPTeX ver.0.8\\ August, 1997}

\title{
Anomalous amplitudes in a thermal bath}

\author{
Robert D. {\sc Pisarski}\footnote{Based upon lectures
presented by R. D. P. at the 
1997 Yukawa International Seminar, YKIS'97, 
``Non-perturbative QCD and the structure of the QCD vacuum'',
Yukawa Institute for Theoretical Physics, Kyoto, 2-12 December, 1997}
T. L. {\sc Trueman}, and M. H. G. {\sc Tytgat}
}

\inst{
Department of Physics\\
Brookhaven National Laboratory\\
Upton, NY 11973, USA
}

\recdate{
April 2, 1996}

\abst{
I review the implications of the axial anomaly in a thermal bath.
I assume that the Adler-Bardeen theorem applies at nonzero temperature,
so that the divergence of the axial current remains is independent
of temperature.
Nevertheless, I argue that while the anomaly doesn't change with
temperature, ``anomalous'' mesonic couplings do.
This is verified by explicit
calculations in a low temperature expansion, and near the
chiral phase transition.
}

\begin{document}

\maketitle

\makeatletter
\if 0\@prtstyle
\def\asp{.3em} \def\bsp{.26em}
\else
\def\asp{.3em} \def\bsp{.3em}
\fi \makeatother

\section{Introduction}

In this paper I provide a pedagogical review of recent work
on the nature of anomalous interactions in a thermal bath.\cite{r1}
To forestall any possible
confusion, at the outset I stress that I assume that
the Adler-Bardeen theorem remains valid in a thermal
bath.  With the proper regularization scheme,
the Adler-Bardeen theorem states that the divergence
of the axial current is given identically by its value
at one loop order.\cite{r2}  Diagramatically, this
is because the axial anomaly is due to the ultraviolet
behavior of one loop graphs.  Since a thermal
bath only affects the infrared, it is natural that
the Adler-Bardeen theorem will still holds at nonzero
temperature.  This is confirmed by explicit calculations.\cite{r3,r4}

Nevertheless, in these notes I show that 
while the anomaly itself remains unchanged, anomalous amplitudes ---
such as multi-point amplitudes between axial vector and vector currents
--- do change in a thermal bath.  This is due directly to the loss of
lorentz covariance in a thermal bath, and is not
special to anomalous currents.  Consider, for example, the two
point function between two conserved, vector currents.
At zero temperature, lorentz invariance and current conservation
implies that this two point function involves only one
scalar function.  At non zero temperature, the lack of explicit
lorentz covariance implies that there are four scalar functions;
imposing current conservation leaves three independent functions.
Similar considerations enter for anomalous currents; there
are some slight changes because the currents are anomalous instead
of conserved, but this is really secondary.

This understanding originated in work by 
Itoyama and Mueller;\cite{r4}
it was then demonstrated by explicit calculations in
various specific models.\cite{r1}  Unlike our
presentation in the literature, in these
notes I begin by reviewing the general analysis,\cite{r1}
and then summarize how these results are realized about
zero temperature, and then about the chiral phase transition.
While it is opposite to how things were understood historically,
the presentation is more logically coherent.

\section{General analysis}

Consider a vector current, 
$J_\alpha$, and an
axial current, $J_{5,\gamma}$.
I assume that the vector current is conserved, 
\begin{equation}
\partial^\alpha J_\alpha = 0 \; ,
\end{equation}
while the axial current is anomalous,
\begin{equation}
\partial^\alpha J_{5,\alpha}
= - \frac{e^2 N_c}{48 \pi^2} F_{\alpha \beta} 
\widetilde{F}^{\alpha \beta} \; .
\label{eq:4.0a}
\end{equation}
The Adler-Bardeen theorem is the statement that
the coefficient of the right hand side, computed
to one loop order, is exact to any loop order;\cite{r2}
this coefficient is also expected to be independent of temperature and
density.\cite{r3,r4}

The classic quantity to compute is the 
three point Green's function between one axial vector current and
two vector currents:
\begin{eqnarray}
\label{eq:4.1}
{\cal T}_{\alpha\beta\gamma}(P_1,P_2;T) 
&=& - i \, e^2 \int d^4 X_1 d^4 X_2\, 
e^{i(P_1\cdot X_1 + P_2\cdot X_2)}\,\\
& \times & 
\frac{\mbox{\rm Tr}\left(e^{-H/T} J_\alpha(X_1) 
J_\beta(X_2) J_{5,\gamma}(0) \right)}{\mbox{\rm Tr}(e^{-H/T})}
\; . \nonumber
\end{eqnarray}
This is a true Green's function in a thermal bath,
with $H$ the hamiltonian.  Given the abelian anomaly, 
this $AVV$ correlation function is the
simplest Green's function in which the anomaly enters.  
For the nonabelian anomaly, besides $AVV$, 
there are also box diagrams, such as $AVVV$, and pentagon
diagrams, such as $AVVVV$ and $AAAVV$.
All of these other Green's functions can be analyzed by similar
means, although because of a proliferation of
independent functions, the details become increasingly complicated.

Since the vector currents are conserved, 
${\cal T}_{\alpha \beta \gamma}$ satisfies
\begin{equation}
\label{eq:4.3}
P_1^\alpha {\cal T}_{\alpha\beta\gamma} 
= P_2^\beta  {\cal T}_{\alpha\beta\gamma} = 0 \; ;
\end{equation}
similarly, the divergence of the axial vector current is anomalous,
\begin{equation}
\label{eq:4.2}
Q^\gamma {\cal T}_{\alpha\beta\gamma} = - {e^2 N_c\over 12 \pi^2}
\,\varepsilon_{\alpha\beta\gamma\delta}\, P_1^\gamma P_2^\delta \; ,
\end{equation}
$Q = P_1 + P_2$.  

I now need to relate this $AVV$ correlation function to the amplitude
for pion decay; to do so, 
I follow Shore and Veneziano.\cite{r5}
At low temperature the pion couples to the axial current as
\begin{equation}
\langle 0\vert J_{5, \alpha}^a\vert \pi^b(Q)\rangle 
= i Q_\alpha f_\pi \delta^{a b} \; .
\label{eq:4.4}
\end{equation}
I work in the chiral limit, in which the pions
are massless.
Strictly speaking, (\ref{eq:4.4}) is valid only to lowest nontrivial
order about zero temperature.  In a nonlinear sigma model
with pion decay constant $f_\pi$, that is $\sim T^2/f_\pi^2$;
to this order, pions obtain a finite, temperature dependent
renormalization constant, but otherwise they propagate without damping.
To higher order, however, such as $\sim T^4/f_\pi^4$, pions
are damped,\cite{r6} and their self energy
acquires an imaginary part even on mass shell.  
This damping in turn implies that there is a nontrivial
matrix element not just between the axial vector current and
one pion, but also between the axial vector current and three pions.
This complicates, but does not invalidate,  the following analysis.
Implicitly, I ignore pion damping, because even then I shall see that 
the direct connection between the axial anomaly
and the electromagnetic decay of the $\pi^0$ is already lost.  

To obtain the amplitude for $\pi^0\rightarrow \gamma \gamma$,
I introduce $Q^2$ times
the matrix element between two vector currents
and a pion, 
\begin{eqnarray}
\label{eq:4.6}
{\cal T}_{\alpha\beta} &=&  e^2 Q^2 \int d^4 X_1 d^4 X_2\, 
e^{i(P_1\cdot X_1 + P_2\cdot X_2)}\\
&\times & 
\frac{\mbox{\rm Tr}\left(e^{-H/T} J_\alpha(X_1) 
J_\beta(X_2) \pi(0)\right)}{\mbox{\rm Tr}\left(e^{-H/T}\right)}
\; . \nonumber
\end{eqnarray}
This is related to the pion decay amplitude as
\begin{equation}
\label{eq:4.7}
{\cal M} = \lim_{Q^2\rightarrow 0}
\epsilon_1^\alpha \epsilon_2^\beta \,
{\cal T}_{\alpha\beta} \; ,
\end{equation}
where $\epsilon_1^\alpha$ and $\epsilon^\beta_2$
are the polarization tensors for the two photons.

The original amplitude contains terms which are one particle
irreducible; in addition, it also contains terms which are one particle
reducible.  Of the latter, I pick out those
which are one pion reducible.  I then 
subtract the one pion pole term from (\ref{eq:4.1})
to define $\widehat {\cal T}_{\alpha\beta\gamma}$,
\begin{equation}
\label{eq:4.5}
\widehat {\cal T}_{\alpha\beta\gamma} = {\cal T}_{\alpha\beta\gamma} +  
f_\pi  \; Q_\gamma \,  { 1\over Q^2} \,
{\cal T}_{\alpha\beta} \; .
\end{equation}
$\widehat {\cal T}_{\alpha\beta\gamma}$, which be definition is
one pion irreducible, satisfies Ward identities similar to those
for ${\cal T}_{\alpha\beta\gamma}$.  The condition for
current conservation is identical,
\begin{equation}
\label{eq:4.9}
P_1^\alpha \widehat {\cal T}_{\alpha\beta\gamma} = P_2^\beta
\widehat {\cal T}_{\alpha\beta\gamma} = 0 \; .
\end{equation}
The anomalous Ward identity differs, receiving a contribution
from the one pion pole,
\begin{eqnarray}
\label{eq:4.8}
Q^\gamma \widehat {\cal T}_{\alpha \beta \gamma} =  f_\pi \;
{\cal T}_{\alpha\beta} - { e^2 N_c\over 12 \pi^2}
\,\varepsilon_{\alpha\beta\gamma\delta}\, P_1^\gamma P_2^\delta \; .
\end{eqnarray}
I now see what general relations can be deduced from these
relations, using Bose symmetry between
the two photons, $P_1, \alpha \rightleftharpoons P_2, \beta$.

I first discuss
zero temperature, where euclidean invariance can be invoked.
The most general pseudo-tensor $\widehat {\cal T}_{\alpha\beta\gamma}$ 
which satisfies all of our conditions can be shown to
involve only three terms:
\begin{eqnarray}
\label{eq:4.10}
\widehat{\cal T}_{\alpha\beta\gamma} &=& T_1\, 
\varepsilon_{\alpha\beta\gamma\delta}( P_1^\delta
- P_2^\delta) \\
&+& T_2 \,(\varepsilon_{\alpha\gamma \delta \kappa} P_{2}^\beta\, - 
\varepsilon_{\beta\gamma \delta \kappa} P_1^\alpha) P_1^\delta
P_2^\kappa \nonumber \\
&+&  T_3\, (\varepsilon_{\alpha\gamma \delta \kappa} P_1^\beta - 
\varepsilon_{\beta\gamma \delta \kappa} P_2^\alpha) 
P_1^\delta P_2^\kappa \; . \nonumber 
\end{eqnarray}
Current conservation, (\ref{eq:4.9}), gives
\begin{equation}
\label{eq:4.12}
T_1 + P_1^2 \, T_2 + P_1\cdot P_2 \, T_3 = 0 \; ,
\end{equation}
while from the anomalous Ward identity, (\ref{eq:4.8}), 
\begin{equation}
\label{eq:4.11}
- 2\, T_1 = f_\pi g_{\pi\gamma\gamma} - {e^2 N_c\over 12 \pi^2} \; .
\end{equation}
Combining these two relations,
\begin{equation}
\label{eq:4.13}
2 P_1^2 \, T_2 + 2 P_1 \cdot P_2  \, T_3
= f_\pi g_{\pi\gamma\gamma} - {e^2 N_c\over 12 \pi^2} \; .
\end{equation}
Putting the photons on their mass shell $P_1^2 = P_2^2$, 
the left hand side in~(\ref{eq:4.13}) reduces to $Q^2 T_3$.
Since by definition I constructed $\widehat{\cal T}$ to
be one pion irreducible, this must vanish on the pion
mass shell, $Q^2 \rightarrow 0$; there is no possibility for
poles in $1/Q^2$ to enter into $T_3$.
Hence the left hand side of (\ref{eq:4.13}) vanishes, and
I obtain a relation between $g_{\pi \gamma \gamma}$
and the coefficient of the axial anomaly,
\begin{equation}
\label{eq:4.14}
0 = f_\pi g_{\pi\gamma\gamma} - {e^2 N_c\over 12 \pi^2} \; .
\end{equation}

This analysis, and especially the tensor
decomposition of (\ref{eq:4.10}), is identical to the 
derivation of the
Sutherland-Veltman theorem.\cite{r7}
Historically, this theorem predated the anomaly, 
and was used originally to conclude that $g_{\pi \gamma \gamma} = 0$;
that is, that the electromagnetic decay of the neutral pion
was chirally suppressed.  
By adding the axial 
anomaly through the anomalous Ward identity of 
(\ref{eq:4.8}), however, I obtain 
$g_{\pi \gamma \gamma} \sim e^2 N_c/f_\pi$, (\ref{eq:4.14}).
This was first derived by Adler,\cite{r2} and is 
reasonably accurate.  
From a modern perspective, then, it is precisely the anomaly
{\it plus} the Sutherland-Veltman theorem which 
allows us to relate the amplitude for $\pi^0 \rightarrow \gamma \gamma$
to the coefficient of the axial anomaly.
If, for example, $Q^2 T_3$ did not vanish, then while there
would be a condition from the anomalous Ward identity, it wouldn't
uniquely predict the amplitude for $\pi^0 \rightarrow \gamma \gamma$.

This is something like what happens at nonzero temperature.
I follow Itoyama and Mueller,\cite{r4}
and write the most general tensor decomposition
for $\widehat {\cal T}_{\alpha \beta \gamma}$.  
The crucial point is obvious: in a thermal bath,
euclidean symmetry is lost, so that I can introduce
a new vector, $n_\mu = (1,\vec{0})$, which denotes the
rest frame of the bath.
There are now many more tensors which can enter.  A
partial list of the new tensors includes
\begin{eqnarray}
\label{eq:4.15}
\widehat{\cal T}_{\alpha\beta\gamma} &=& T_1\, 
\varepsilon_{\alpha\beta\gamma\delta}( P_1^\delta - P_2^\delta) \\
&& \mbox{} + T_2 \,(\varepsilon_{\alpha\gamma\delta\kappa} \, P_2^\beta - 
\varepsilon_{\beta\gamma\delta\kappa}\, P_1^\alpha) P_1^\delta
P_2^\kappa \nonumber \\
&& \mbox{} + T_3\, (\varepsilon_{\alpha\gamma\delta\kappa} \, P_1^\beta - 
\varepsilon_{\beta\gamma\delta\kappa}\, P_2^\alpha ) P_1^\delta P_2^\kappa
\nonumber\\
&& \mbox{} + T_4 \, n\cdot Q \, 
\varepsilon_{\alpha\beta\delta\kappa} \, P_1^\delta P_2^\kappa \,
n^\gamma\nonumber\\
& & \mbox{} + T_5 
(n\cdot P_2 \, \varepsilon_{\alpha\gamma\delta\kappa} \, n^\beta  -
n\cdot P_1 \, \varepsilon_{\beta\gamma\delta\kappa} \, n^\alpha ) 
P_1^\delta P_2^\kappa\nonumber
\\
&& \mbox{} + \ldots\nonumber
\end{eqnarray}
I have only included the terms in $\widehat{\cal T}$ which
contribute to the Ward identities of (\ref{eq:4.9}) and 
(\ref{eq:4.8}); what other tensors enter will not matter for
our considerations.  Current conservation gives
\begin{equation}
\label{eq:4.17}
T_1 + P_1^2 \;T_2 + P_1 \cdot P_2\; T_3 
+ (n\cdot P_1)^2 \; T_5 = 0 \; ,
\end{equation}
while the anomalous Ward identity fixes

\begin{equation}
\label{eq:4.16}
- 2 T_1 + (n\cdot Q)^2 \, T_4  =  f_\pi(T) g_{\pi\gamma\gamma}(T) -  
{e^2 N_c\over 12 \pi^2} \; .
\end{equation}
Notice that at nonzero temperature, I allow $f_\pi$ and
$g_{\pi\gamma\gamma}$ to depend upon temperature; explicit
calculation shows that they can and do.
Combining these two relations, I find 
$$
2 P_1^2 \, T_2 + 2 P_1 \cdot P_2  \, T_3
+(n\cdot Q)^2 \, T_4 + 2 (n \cdot P_1)^2 \, T_5
$$
\begin{equation}
= f_\pi(T) \, g_{\pi\gamma\gamma}(T) -  {e^2 N_c\over 12 \pi^2} \; .
\label{eq:4.18a}
\end{equation}
This relation is valid for arbitrary momenta.  I then
put the photons on their mass shell,
$P_1^2 = P_2^2 = 0$, 
as well as the pion, $Q^2 \rightarrow 0$.  Since
by construction $T_3$ is one pion irreducible, it cannot have
a pole in $\sim 1/Q^2$, and so I find
\begin{equation}
\label{eq:4.18}
(n\cdot Q)^2\, T_4 + 2 (n\cdot P_1)^2\, T_5   
= f_\pi(T) \, g_{\pi\gamma\gamma}(T) -  {e^2 N_c\over 12 \pi^2} \; .
\end{equation}
The terms on the right hand side are as at zero temperature.
But now I find two terms on the left hand side, which involve
the energy squared for the pion, $(n\cdot Q)^2$, and the same for
one photon, $n \cdot P_1$.  Even letting all fields go on their
mass shell, I can do so without letting the energies vanish.
Further, there is no reason why the amplitudes $T_4$ and $T_5$ 
should vanish at these point.  Thus I see that at nonzero
temperature, while there is a condition from the axial anomaly,
it cannot be used to uniquely relate 
$g_{\pi \gamma \gamma}(T)$, $f_\pi(T)$, and the coefficient of
the axial anomaly; I also need the values of $T_4$ and $T_5$.

In general terms, I have assumed that the Adler-Bardeen theorem
applies.  What failed was the Sutherland-Veltman theorem: the
terms on the left hand side of (\ref{eq:4.18}) don't need to,
and in general don't, vanish.  

What happens beyond lowest order in an expansion about
zero temperature?  When the effects of pion damping 
are included, one will have to deal with a pion mass shell
which is not only off the light cone, but also has an imaginary
part.  Since Goldstone's theorem remains valid in a thermal
bath, this is not a problem in principle.  More
states contribute to the anomalous Ward identity, but
it remains valid.

Initially, one might well wonder why the Sutherland-Veltman
theorem should apply at zero, but not at any nonzero, temperature.
Even at zero temperature, though, it should be remembered
that the Sutherland-Veltman theorem only applies in a very
strictly defined regime: in the chiral limit, with all fields,
the pion and both photons, on their mass shell.
For example, consider the case
in which one is in the chiral limit, but only one photon is
on its mass shell.  Then the left hand side of 
hand side of (\ref{eq:4.18a}) doesn't vanish, 
and $g_{\pi \gamma \gamma}$ is not simply related to the anomaly.
In fact, consider the limit in which $P_1^2$ is large;
then even in the chiral limit, $Q^2 \rightarrow 0$,
$g_{\pi \gamma \gamma} \sim e^2 (f_\pi/P_1^2)$.\cite{r8}
This agrees with a simple power counting of the underlying quark diagrams
which contribute to $g_{\pi \gamma \gamma}$: any coupling
falls off like powers of momenta at large momenta.  

\section{Low temperature}

About zero temperature, it is most convenient to use a nonlinear
sigma model.  This is a nonrenormalizable theory, but one
can still treat it with a cutoff. Furthermore, for the
temperature dependent effects which I are interested in,
everything is obviously ultraviolet finite.

The technical details of computing with a nonlinear sigma
model are involved.  To include anomalous couplings,
one adds a Wess-Zumino-Witten term to the (gauged)
nonlinear sigma model lagrangian, and then compute
loop effects with that lagrangian.  For definiteness, all
calculations are for two massless flavors of quarks.

At zero temperature, one can perform a nontrivial check
of both the Sutherland-Veltman and Adler-Bardeen theorems.
At tree level, the Wess-Zumino-Witten term defines a coupling
between the pion and two photons, which satisfies
\begin{equation}
f_\pi g_{\pi \gamma \gamma} = \frac{e^2}{4 \pi^2} \; .
\label{da}
\end{equation}
This is a relationship between bare quantities in the tree
lagrangian.  One can then compute to one loop order.  Since
since the nonlinear sigma model is nonrenormalizable, it
is unremarkable to find that 
quadratic divergences arise.
For example, the relationship between the bare and
renormalized pion decay constants is
\begin{equation}
f^{ren}_\pi = 
\left ( 1 - 
\frac{1}{f_\pi^2} \int {d^4\!K\over (2 \pi)^4} \; {1\over K^2} 
\right) f_\pi \; .
\label{db}
\end{equation}
The renormalized coupling between a
pion and two photons is similarly found to be
\begin{equation}
g^{ren}_{\pi \gamma \gamma} =
\left ( 1 +
\frac{1}{f_\pi^2} \int {d^4\!K\over (2 \pi)^4} \; {1\over K^2} 
\right) g_{\pi \gamma \gamma}  \; .
\label{dc}
\end{equation}
Now of course to be well defined, I should introduce some
(chirally invariant) regularization scheme, such as dimensional
regularization.  But in fact no matter how one regulates
the quadratic divergence in these expressions, it is clear that to 
one loop order, $\sim 1/f_\pi^2$, 
the product of renormalized quantities satisfies 
 the same relationship as in~(\ref{da})~\cite{r10},
\begin{equation}
f^{ren}_\pi g^{ren}_{\pi \gamma \gamma} = 
f_\pi g_{\pi \gamma \gamma} = \frac{e^2}{4 \pi^2} \; .
\label{dd}
\end{equation}
In terms of the previous
section, no diagrams contribute to $T_2$ or
$T_3$.  Two diagrams contribute
to $T_1$, but cancel against each other.  Because
$T_1 = T_2 = T_3 =0$, at zero temperature
both the Sutherland-Veltman and Adler-Bardeen theorems apply.  

At nonzero temperature, at first one might reason\cite{r11}
that topology should similarly constrain the couplings
as in (\ref{dd}).  Indeed, the diagrams are absolutely identical
to those at zero temperature.  Thus one would expect that
I simply extract the temperature dependent piece from
the (quadratic) divergences in $f_\pi$ and $g_{\pi \gamma \gamma}$,
as 
\begin{equation}
\int {d^4\!K\over (2 \pi)^4} \; {1\over K^2}
= (T=0) + {T^2 \over 12} \; .
\label{de}
\end{equation}
However the divergence at zero temperature is regulated, the
temperature dependent piece in the integral is perfectly
well defined.

This works for the pion decay constant; to $\sim T^2/f_\pi^2$,
it decreases as this naive argument and (\ref{db}) would suggest,
\begin{equation}
f_\pi(T) = \left(1 - {1 \over 12} 
{T^2\over f_\pi^2}\right) f_\pi \; .
\label{df}
\end{equation}
The only diagrams which contribute to $f_\pi$, however,
are wave function renormalization for the pion, and a
renormalization of the axial vector current.  Both of
these diagrams are ``tadpole'' type diagrams, and are
clearly independent of the external momentum.

This is not true for the coupling between a pion and two
photons.  Most of the diagrams which contribute are tadpole
type diagrams, but one, in which a single photon couples
to a pion loop, is not.  For this last diagram, the momentum
dependence must be treated with care.  Doing so, one finds
the following.  To lowest order about zero temperature,
$f_\pi(T)$ decreases, (\ref{df}).  If the result at
nonzero temperature were like that at zero temperature,
(\ref{dc}), then (\ref{de}) would predict that 
$g_{\pi \gamma \gamma}$ increases with temperature.  Instead,
I find that it decreases,
\begin{equation}
g_{\pi \gamma \gamma}(T) 
= \left(1 - {1 \over 12} {T^2 \over f_\pi^2}\right) 
\; g_{\pi \gamma \gamma} \; .
\label{dg}
\end{equation}

In terms of the anomalous Ward identity, calculation shows
that at nonzero temperature, while no diagrams contribute to
$T_2=T_3=0$, $T_1$ is nonzero:
\begin{equation}
T_1 = \frac{T^2}{12 f_\pi^2} \;
{e^2  \over 4 \pi^2} \; .
\label{dga}
\end{equation}
For the terms special to nonzero temperature, $T_4$ vanishes,
while 
\begin{equation}
T_5 =  - { 1\over (n \cdot P_1 )^2} 
\; \frac{T^2}{12 f_\pi^2} \;
{e^2  \over 4 \pi^2} \; .
\label{dgb}
\end{equation}
Comparing to the left hand side of (\ref{eq:4.18}), because
$T_5$ is nonzero, 
the Sutherland-Veltman theorem does not apply even to
leading order at nonzero
temperature, $\sim T^2/f_\pi^2$.  In contrast, to this order
the anomalous Ward identity, and so
the Adler-Bardeen theorem, are satisfied.  In fact,
$T_5$ and the anomalous Ward identity provides a nice check of our
results for $f_\pi(T)$ and $g_{\pi \gamma \gamma}(T)$
in (\ref{df}) and (\ref{dg}).

As the Sutherland-Veltman theorem fails even when it
pion damping can be neglected, there is no point in considering
it to higher order in an expansion about zero temperature.  
Surely the Adler-Bardeen theorem remains valid, although how in detail
it is manifested is presumably involved and a question of interest.

Diagramatically, while the same diagrams 
contribute to $g_{\pi \gamma \gamma}(T)$ to one loop order
at $T=0$ and to $\sim T^2/f_\pi^2$, it is the delicate
momentum dependence of one diagram at nonzero temperature which 
gives rise to the unexpected result in (\ref{dg}).  
This comes about because of a surprising analogy.  The 
momentum dependent diagram which
contributes to $g_{\pi \gamma \gamma}$ is proportional to 
\begin{equation}
T \sum_{n = - \infty}^{+ \infty} \int {d^3\!k\over (2\pi)^3}
 {K^\alpha K^\beta \over K^2 (K-P)^2} \; ,
\label{dh}
\end{equation}
where $P=(p^0,\vec{p})$ is an external momentum for the gauge field.

Exactly the same function enters into what appears to be
a very different problem: the self energy for a gauge field,
coupled to massless fields, in the limit of 
{\it high}, as opposed to low, temperature.  It is well
known for gauge fields that the self energy is an involved
function of the external momenta, and that different results
are obtained depending upon how the zero momentum limit
is reached: in particular, the static limit,
$p^0=0$, then $\vec{p}\rightarrow 0$, differs from the limit
on the light one, 
$p^0 = i \omega$, $\omega = p \rightarrow 0$.
Technically, this
is why the guess for $g_{\pi \gamma \gamma}(T)$ failed:
since $P$ is the
momentum of the external photon,
the zero momentum limit which enters isn't the static
one, but that on the light cone. 

This dependence on the external momentum is more easily
understood when one constructs an effective lagrangian
for $\pi^0 \rightarrow \gamma \gamma$:
\begin{equation}
{\cal L}_{\pi^0 \gamma \gamma}(T) 
=  \left( {e^2 N_c\over 48 \pi^2} \right)  \; {1 \over f_\pi(T)}\;
\pi^0 F_{\alpha \beta} \widetilde{F}^{\alpha \beta}
\label{di}
\end{equation}
$$
 - \frac{T^2}{12 f_\pi^2} \left( {e^2 N_c\over 48 \pi^2} \right)
\int \frac{d \Omega_{ \hat{k} } }{4 \pi} \; 
H_{\gamma \alpha}
\frac{ \hat{K}^\alpha \hat{K}^\beta}{- (\partial \cdot \hat{K})^2}
F_{\gamma \beta} \; ,
$$
where 
$\widetilde{F}^{\alpha \beta} =
\epsilon^{\alpha \beta \gamma \delta} F_{\gamma \delta}/2$, 
$
H_{\alpha\beta} = \partial_\alpha H_\beta - \partial_\alpha H_\beta 
$,
and
\begin{equation}
H_\alpha = {1\over f_\pi}\varepsilon_{\alpha\beta\gamma\delta} 
F_{\beta\gamma} \partial_\delta \pi^0 \; .
\label{dk}
\end{equation}
The vector $\hat K = ( i, \hat{k})$; one then integrates
over all angles $\hat{k}$.  This integration represents
the hard, massless field in the one loop integral.

The important aspect of (\ref{di}) is that it is nonlocal.
The nonlocality is familiar from hard thermal loops in gauge
theories, and is responsible for the sensitivity to how the
zero momentum limit is reached.  The complete expression for
the temperature dependent terms in the Wess-Zumino-Witten
lagrangian was derived by Manuel.\cite{r1}

\section{Near the chiral phase transition}

In this section I explain what was historically the first
example of how the coupling of 
$\pi^0 \rightarrow \gamma \gamma$ changes with
temperature.\cite{r1}  I work near the chiral phase transition,
which is assumed to be of second order, and show that
in this limit, $g_{\pi \gamma \gamma}$ vanishes as
the point of phase transition is approached.  I emphasize
the simplicity of the phenomenon; for technical reasons
which I will discuss, the analysis is not as complete
as about zero temperature.

I employ a constituent quark model.  At the outset I confess
that I don't think that this model is at all a realistic
model to calculate detailed properties of the chiral phase
transition.  I do think it is good enough to explain
the qualitative physics, in essence at the level
of a type of mean field theory.

The coupling between the mesons and quark fields is
take to be
\begin{equation}
{\cal L} =
\overline{\psi} \left( \not \!\! D + 2 \widetilde{g} 
\left( \sigma t_0 + i 
\vec{\pi} \cdot \vec{t} \gamma^5 \right)
\right) \psi \; .
\label{ea}
\end{equation}
I take two flavors, with $t_0 = {\bf 1}/2$, and 
$tr(t^a t^b) = \delta^{a b}/2$.  This lagrangian
is invariant under the standard chiral symmetry
of $SU(2)_\ell \times SU(2)_r$.  The meson fields
include the $\sigma$ meson and pions.

At zero temperature I assume that the $\sigma$ field acquires
a vacuum expectation value; for two flavors, at tree level
the pion decay constant is identically this v.e.v,
$f_\pi = \sigma_0$.  (The ratio of $f_\pi/\sigma_0 \neq 1$
for three or more flavors).  At nonzero temperature,
$f_\pi$ is no longer strictly equal to $\sigma_0$; this
can be seen by an expansion about zero temperature,
as the terms $\sim T^2/f_\pi^2$ differ.  
If $T_{ch}$ is the temperature for the chiral
phase transition, which is assumed to be of second order, then
both should vanish 
as $T \rightarrow T_{ch}^-$ in the same manner,
$\sigma_0 \sim f_\pi \sim (T_{ch} - T)^{\beta}$.
In mean field theory, $\beta = 1/2$.

The coupling between a pion and two photons is given by
a triangle diagram, similar to that which contributes to the 
axial anomaly.  For the axial anomaly, however, one 
computes the divergence of the axial current; in the fermion loop, the
vertex 
brings in one factor of the momentum, since it is the divergence
I are computing, and one factor
of the Dirac matrix $\gamma^5$, as an axial current.  For the coupling
between a pion and two photons, the pion vertex 
brings in one $\gamma^5$,
but no factor of the momentum, just the coupling $\widetilde{g}$.
For the divergence of
the axial anomaly, the factor of the momentum means that
the triangle diagram is sensitive to the ultraviolet regime,
and completely insensitive to the infrared regime.  For
the decay of a pion, without the power of momentum upstairs,
the associated triangle diagram becomes completely
insensitive to the ultraviolet, but sensitive to the infrared.

For the axial anomaly, the diagram involves the following
trace over Dirac matrices:
\begin{equation}
tr(\gamma^\delta \gamma^5 \Delta(K) 
\gamma^\alpha \Delta(K-P_1) \gamma^\beta \Delta(K-P_2)) \; ,
\label{eb}
\end{equation}
with $\Delta$ the fermion propagator.  The detailed momentum
dependence doesn't matter for my arguments.  As stated,
for the axial anomaly only the ultraviolet behavior matters,
so I can take the propagators to be massless.  Then I
are left with the trace of $\gamma^5$ times six Dirac
matrices; since the trace of $\gamma^5$ times four, six,
etc. Dirac matrices is nonzero, I find a nontrivial result.

In contrast, for pion decay the trace over Dirac matrices is
\begin{equation}
tr(\gamma^5 \Delta(K) 
\gamma^\alpha \Delta(K-P_1) \gamma^\beta \Delta(K-P_2)) \; .
\label{ec}
\end{equation}
If I take each fermion propagator to be massless, the
integral vanishes identically, since I have the trace
of $\gamma^5$ times five Dirac matrices.  To have
any nonzero result, there must be one power of the 
mass from a fermion propagator.  This is the essential
origin of the suppression of pion decay near the chiral
phase transition.

The overall behavior of the diagrams can be estimated
on the basis of power counting and gauge invariance.
Gauge invariance tells us that the gauge fields have
to enter in through the form $F_{\alpha \beta} \widetilde F^{\alpha \beta}$.
Thus the divergence of the axial current, and this operator,
are each dimension four, so the coefficient is a pure number.
The Dirac trace in (\ref{eb}) should just give us that part of
the one loop result.

For the coupling between a pion and two gauge fields, since the
latter enter as $F_{\alpha \beta} \widetilde F^{\alpha \beta}$,
the coupling must have dimensions of inverse mass.  
One can read off the relevant factors without direct computation:
there is one factor of $\widetilde{g}$ from the coupling,
and one factor of $m$ from the Dirac trace in (\ref{ec}).
That leaves an integral with dimensions of mass squared; about
zero momentum, the only natural mass scale is $1/m^2$.  Thus
at zero temperature,
\begin{equation}
\sim e^2 \; \frac{\widetilde{g} \, m}{m^2} \; \pi^0 \; 
F_{\alpha \beta} \widetilde F^{\alpha \beta}
= \frac{e^2}{12 \pi^2 }\; \frac{1}{f_\pi} \; \pi^0 \; 
F_{\alpha \beta} \widetilde F^{\alpha \beta} \; .
\label{ed}
\end{equation}
Here I have used the fact that as can be read off from
(\ref{ea}), the constituent quark
mass $m = \widetilde{g} \sigma_0 = \widetilde{g} f_\pi$.
The result in (\ref{ed}) is the first term in the Wess-Zumino-Witten
lagrangian.  The full lagrangian is complicated, although
the overall coefficient is dictated by topology, which indirectly
reflects the topology underlying the Adler-Bardeen theorem.

To estimate the coupling at nonzero temperature, I only need
recognize that while the factors of $\widetilde{g}$
and $m = \widetilde{g} f_\pi$ remain the same, the integral,
being sensitive to the infrared, changes.  
The diagram involves fermions at nonzero temperature.  In
the limit in which 
the constitutent quark mass is much less than the 
temperature, the only mass scale in the loop integral is the
tempeature $T$.
Due to Fermi-Dirac statistics, there are no infrared divergences,
and the temperature just acts as an infrared cutoff, as
the coupling between $\pi \rightarrow \gamma \gamma$ becomes
\begin{equation}
\sim e^2 \; \frac{\widetilde{g} \, m(T)}{T^2} \; \pi^0 \;
F_{\alpha \beta} \widetilde F^{\alpha \beta} 
= \frac{7 \, \zeta(3) \, e^2 \, \widetilde{g}^2 }{8 \pi^4 T^2} \; 
f_\pi(T) \; \pi^0 \; F_{\alpha \beta} \widetilde F^{\alpha \beta} \; .
\label{ee}
\end{equation}
This result is confirmed by direct calculation, which also
gives the coefficient of (\ref{ee}) to one loop order.
I have assumed that the constituent quark mass, $m(T)=
\widetilde{g} f_\pi(T)$, 
changes with temperature, but neglected the dependence
of the coupling constant $\widetilde{g}$ with temperature.
This is because $f_\pi(T)$ changes like a power of $T$,
while as a typical dimensionless coupling constant, $\widetilde{g}$
should only change logarithmically.  Even if the coupling
$\widetilde{g}$ does flow to a fixed point, there is every
reason to believe that it will be a nontrivial fixed point;
then I just replace the ``bare'' $\widetilde{g}$ with
that value.

There is a technical qualification:
the expression in (\ref{ee}) is correct {\it only} at zero momentum,
approached in the static limit.  It is rather more difficult
to compute the analogous amplitudes away from the static limit,
which is the limit of interest of compute the anomalous Ward
identity, with photons and pions which are on their mass shell.
Thus I cannot directly verify how the anomalous Ward identity
is satisfied.  Even with explicit calculation, though, I
can assume that like $g_{\pi \gamma \gamma}$, that
$T_1$, $T_2$, and $T_3$ vanish at $T= T_{ch}$.  
Nevertheless, the anomalous Ward identity can easily be
satisfied if $T_4$ and/or $T_5$ are nonzero.

Besides verifying the anomalous Ward identity, there is another
reason for computing anomalous processes near the chiral phase
transition.  The processes in this section represent one
term in what the Wess-Zumino-Witten lagrangian becomes
near $T_{ch}$.  Like the terms about zero temperature, 
(\ref{dk}), it is undoubtedly nonlocal.  The nonlocality will
not be identical to those of hard thermal loops, though,
but represent a new, nonlocal lagrangian which governs anomalous
processes about the chiral phase transition.

\end{document}